# THE NETWORK ANALYSIS OF URBAN STREETS:
# A PRIMAL APPROACH

Sergio Porta[a], Paolo Crucitti[b], Vito Latora[c]

[a] Dipartimento di Progettazione dell'Architettura, Politecnico di Milano, Italy, sergio.porta@polimi.it
[b] Scuola Superiore di Catania, Italy, pacrucitti@ssc.unict.it
[c] Dipartimento di Fisica e Astronomia, Università di Catania
and INFN Sezione di Catania, Italy, vito.latora@ct.infn.it

## Abstract

The network metaphor in the analysis of urban and territorial cases has a long tradition especially in transportation/land-use planning and economic geography. More recently, urban design has brought its contribution by means of the "space syntax" methodology. All these approaches – though under different terms like "accessibility", "proximity", "integration" "connectivity", "cost" or "effort" – focus on the idea that some places (or streets) are more important than others because they are more central. The study of centrality in complex systems, however, originated in other scientific areas, namely in structural sociology, well before its use in urban studies; moreover, as a structural property of the system, centrality has never been extensively investigated metrically in geographic networks as it has been topologically in a wide range of other relational networks like social, biological or technological.

After two previous works on some structural properties of the dual and primal graph representations of urban street networks (Porta et al. 2004; Crucitti et al. 2005), in this paper we provide an in-depth investigation of centrality in the primal approach as compared to the dual one, with a special focus on potentials for urban design. An innovative methodology for the analysis of geographic spatial networks, which we term Multiple Centrality Assessment (MCA), is defined and experimented on four real urban street systems. MCA, it turns out, provides a new perspective to the network analysis of spatial systems, which is inherently different from space syntax in that: 1. it is based on a primal, rather than a dual, graph representation of street patterns; 2. it works within a fully metric, rather than topologic, framework; 3. it investigates a set of peer centrality indices rather than just a principal one. We show that spatially, some centrality indices nicely capture the "skeleton" of the urban structure that so much impacts on spatial cognition and collective behaviors; we also show that statistically, centrality distributions consistently characterize geographically different patterns, and power laws emerge in self-organized cities.

## 1 Introduction: which order for urban street patterns?

«There is often not as much perfection in works composed of several pieces and made by the hands of diverse masters as in those on which one individual alone has worked. […] Thus those ancient cities which, having been, in the beginning, only villages, have become, with the passage of time, great towns are ordinarily so badly put together, compared with those regular places which an engineer traces out on a plain according to his fantasy, that, even though, considering each one of their edifices separately, one often finds in them as much art as, or more of it than, in those of the others, nonetheless, to look at how they are arranged, a big one here, a small one there, and how they render the streets crooked and unequal, one would say that it is chance, rather than the will of a certain man using reason, that has placed them thus» (Descartes, 1994 c.1637, p.27). At the dawn of modernity, Descartes firmly defined the Euclidean geometry as the sole kind of order acknowledgeable by the eyes of the rational man while dealing with the form of environmental configurations, like gardens, landscapes, streets and cities. Some three centuries later, Le Corbusier brutally reacted against Sitte's recall to the social aesthetics of medieval windy patterns with his well known invective: «La rue courbe est le chemin des ânes, la rue droite le chemin des Hommes» («The windy street is the path of asses, the straight street is the path of men», Le Corbusier, 1994 c.1925). Only asses, The Master went on, could have designed the historical cities of Europe, with all that mess of narrow streets and that horrible, *chaotic* puzzle of intersections and squares. Still in our days, the power of Euclidean geometry is immensely influential for architects and urban designers, almost an axiom when it comes to the design of streets, towns and cities. Not unlike the urban renewal good old days, still today old neighborhoods are often underestimated in their most fundamental values: well, yes, they might be considered picturesque, even attractive due to some unique architectural features, but their structure is not that valuable: it is *disordered*.

Against this modernist stigmatization, a whole stream of counter-arguments have been raised up since the early Sixties in the name of the "magic" of old cities (Jacobs, 1993). The claim was not just about aesthetics: it was about livability. The modern city, so it goes, is hard to live in. The social success of an urban settlement, which is conceived as the unpredictable outcome of safety, trust, economic vitality and diversity, sprouts from the complex, unplanned interactions of countless different routes and experiences in a suitable environment. Warren Weaver, quoted in Jane Jacobs (Jacobs, 1961), argued that science is not just one (Weaver, 1948). To be true, there are many of them. Moreover, science evolves in history. Thus, simple problems have been investigated by a different kind of science than complex-disorganized ones, and complex-organized problems require a different science again. Well – Jane Jacobs said – the point with "orthodox" planners is that they used the right tools for the wrong problem: cities – she argued – are complex-organized problems and as such, in order to be understood, require to be approached with a new science, the science of complexity. «Under the seeming disorder of the old city, wherever the old city is working successfully, is a marvelous order for maintaining the safety of the streets and the freedom of the city. It is a complex order» (Jacobs, 1961, p.50): only by means of the new science of complexity the "marvelous" complex order of the old city can be revealed, an order which, unlike the Euclidean geometry, is not visible at a first glance, is not imposed by any central agency, rather is the result of the fine-grained, gradual contribution of countless agents in time, each following his/her or its own trajectory. That order, Jane Jacobs concluded, is an organic one; it is the order of life, the only one that can contribute to the actual livability of neighborhoods and cities, thus the one that should drive us to the sustainable city of the future (Newman and Kenworthy, 1999).

Everyone knows, in the network community, an area of scientific research that investigates complex systems through the use of the metaphor of networks and the mathematics of graphs, that a good deal of achievements have been gained in the very last years since the seminal work of Watts and Strogatz on the so-called "small worlds" in 1998 (Watts and Strogatz, 1998). The availability of detailed maps of a broad range of self-organized systems – ranging from natural to man-made, from chemical to biological to neural, sexual and even linguistic systems and many others again – as well as an immensely increased computational power, have allowed to understand that all those networks do share some astonishingly similar topological properties (Barabasi, 2002; Albert and Barabasi, 2002). Among others, our studies on *urban street networks* (Porta et al. 2004, Crucitti et al. 2005) – a kind of geographic networks, which are a specific family of complex systems characterized by planarity and metric distance – have shown that the same properties actually apply to those cases as well. These achievements allow to acknowledge, under the "seeming disorder" of self-organized cities, the clues of a hidden order that operates embedded in the most diverse climatic, geographic, economic, social and cultural conditions, an order shared with most non-geographic natural, biological and social systems (Portugali, 2000; Salingaros, 2003). The "marvelous complex order" of Jane Jacobs' old city, the "magic" of Allan Jacobs' "great streets", now appear a little bit less obscure, as we can clearly see behind the façade the universal sign of nature, the sign of organic complex systems naturally evolved following a rule of preferential attachment and hierarchical topology (Albert and Barabasi, 2002; Ravasz and Barabasi, 2003). The presence of that *kind* of order, an order that should now be regarded as a value, a treasure, something like the "genius of the city" (Whyte, 1988), qualifies – Warren Weaver would say – traditional, self-organized, incremental urban fabrics as a "complex-organized problem": how far is the rich, diverse, organic, vital, deeply ordered structure of self-organized cities from that kind of chaotic mess that, according to the ingenuous eyes of The Master himself, only asses could have designed!

That is why the "new science of networks" is so relevant for urban studies today: because the recognition of the hidden order of self-organized cities is a contribution to the actual overcoming of the modernist heritage in city planning and design, as well as to a new set of goals and opportunities for urban designers, academics and professionals in the field. In this article we make a step forward in this direction with the identification and experimentation of a new set of tools for the primal analysis of *centralities* on urban street systems. In section 2, a short review of centrality indices since the early Fifties to our days is presented; a comparison is then addressed between "space syntax", a well known methodology for the dual analysis of street systems, and previously defined indices of centrality, which leads to understand space syntax under the light of a broader framework and to acknowledge its historical roots. In section 3, a brief discussion of the two different approaches – the primal and the dual – to the graph representation of urban street systems is presented: special attention is posed to the differences between topologic and geographic distance as well as the impact of the implementation of a generalization model on the closeness/integration centrality distribution over a street network. In section 4, selected indices of centrality are investigated over four exemplary real cases of urban street networks in two ways: graphically, by the presentation of thematic maps, and statistically, by plotting their cumulative distributions; both kinds of investigations are implemented over the primal as well as the dual representation of each case. The main message of this article is then presented in section 5 (conclusions): no single-index spatial analysis can depict the whole picture: Multiple Centrality Assessment (MCA), an innovative process of spatial analysis grounded on a set of different centrality indices investigated over a truly primal, metric representation of street networks, can better contribute to an extended comprehension of the "hidden orders" that underlie the structure of real, geographic spatial systems (actually not limited to those of streets and intersections).

## 2 From structural sociology to space syntax: defining centrality indices

A classic in structural sociology, Freeman's master-works on centrality (Freeman, 1977, 1979) reviewed and coordinated under the same roof previous researches addressed since the early fifties (Bavelas, 1948, 1950; Leavitt, 1951; Shimbel, 1953; Shaw 1954, 1964), and defined a first set of indices: degree ($C^D$), closeness ($C^C$) and betweenness ($C^B$) centrality. The basic idea in structural sociology is to represent a group of people in a social or organizational setting as a network whose nodes are the individuals and whose edges are relationships between individuals (Wasserman and Faust, 1994). Bavelas, was the first to realize that central individuals in a social network very often play a prominent role in the group, or in other words a good location in the network structure corresponds to power in terms of independence, influence an control on the others (Bavelas, 1948).

Quite recently, a new impulse to the use of the network metaphor in the scientific community has come from the discovery that complex networks in many different economic, social, natural and man-made systems do share some common structural properties. A first shared property is related to distance and clustering: in fact, it has been shown that most of those networks exhibit the "small-world" property, meaning that the average topological distance between couples of nodes is small compared to the size of the network (it increases only logarithmically with the system's size), despite the fact that the network exhibits a large local clustering typical of regular lattices (Watts and Strogatz, 1998). A second shared property is more related to centrality, and in particular has to do with the node's degree distribution. The degree $k$ of a node is defined as the number of its connections and, as we will see in the following, it is nothing else than a node's centrality measure, $C^D$. The study of a large number of complex systems, including networks as diverse as man-made like World Wide Web and the Internet (Pastor-Satorras and Vespignani, 2004), social like the movie actors collaboration network or networks of sexual contacts (Liljeros et al. 2001), and many biological (Albert and Barabasi, 2002), has shown that, in most of them the degree distribution follows, for large degree $k$, a power law scaling:

$$P(k) \sim N(k) \sim k^{-\gamma} \qquad (5)$$

with the exponent γ being between 2 and 3. In this formula, $N(k)$ is the number of nodes having $k$ links, and $P(k)$ is $N(k)$ divided by the total number of nodes in the network. Networks with such a degree distribution have been named scale-free (Albert and Barabasi, 2002). The results found are particularly interesting because in contrast with what expected for random graphs (Erdös and Rényi, 1959). In fact, a random graph with N nodes and K edges (an average of $\bar{k}$ per node), i.e. a graph obtained by randomly selecting the K couples of nodes to be the connected, exhibits a Poisson degree distribution centred at $\bar{k}$, with an exponential and not a power law behaviour for large values of $k$.

In formal terms, a network can be represented as a graph $G=(N, K)$, a mathematical entity defined by a pair of sets, $N$ and $K$. The first set, $N$, is a nonempty set of $N$ elements called *nodes* or *points*, while $K$ is a set of $K$ elements containing unordered pairs of different nodes called *links* or *edges*. In the following a node will be referred to by its order $i$ in the set $N$ ($1 \leq i \leq N$). If there is an edge between nodes $i$ and $j$, the edge being indicated as $(i,j)$, the two nodes are said to be adjacent or connected. Some times it is useful to consider a valued, or weighted graph $G=(N, K, \Omega)$, defined by three pairs of sets $N$, $K$ and $\Omega$. The set $\Omega$ is a set of $K$ elements being the numerical values attached to the edges, and measuring the strengths of the tight. A graph $G=(N, K)$ can be described by a single matrix, the so-called adjacency matrix $A=\{a_{ij}\}$, a NxN square matrix whose element $a_{ij}$ is equal to *1* if $(i,j)$ belongs to $K$, and zero otherwise. A weighted graph $G=(N, K, \Omega)$ is defined by giving two matrices, the adjacency matrix $A$, defined as above, and a matrix $W$ containing the edge weights. In our particular case, we find more useful to work with lengths in place of weights, so that, instead of the weights matrix $W$ we will consider the lengths matrix $L=\{l_{ij}\}$, a $N \times N$ matrix whose entry $l_{ij}$ is the metric length of the street connecting $i$ and $j$, i.e. a quantity inversely proportional to the weight associated to the edge. In a valued graph, the shortest path length $d_{ij}$ between $i$ and $j$ is defined as the smallest sum of the edges lengths throughout all the possible paths in the graph from $i$ to $j$, while in a non-valued graph it is simply given by smallest number of steps in order to go from $i$ to $j$.

The characteristic path length $L$ (Watts and Strogatz, 1998), is defined as the average length of the shortest paths calculated over all the couples of nodes in the network:

$$L = \frac{1}{N(N-1)} \sum_{i \neq j \in N} d_{ij}. \qquad (1)$$

is a good measure of the connectivity properties of the network. However this index is not well defined for non-connected graphs, unless we make the artificial assumption of a finite value for $d_{ij}$ also when there is no path connecting nodes $i$ and $j$. To overcome this problem a new index of performance has been defined: the global efficiency $E_{glob}$ (Latora and Marchiori, 2001). As the characteristic path length, $E_{glob}$ is a measure of how well the nodes communicate over the network and is defined based on the efficiency in the communication between couples of nodes. The efficiency $\varepsilon_{ij}$ in the communication between two generic nodes of the graph $i$ and $j$, is assumed to be inversely proportional to the shortest path length, i.e. $\varepsilon_{ij}=1/d_{ij}$. In the case $G$ is non-connected and there is no path linking $i$ and $j$ it is assumed $d_{ij}=+\infty$ and, consistently, $\varepsilon_{ij}=0$. The global efficiency of a graph $G$ is defined as the average of $\varepsilon_{ij}$ over all the couples of nodes:

$$E_{glob}(G) = \frac{1}{N \cdot (N-1)} \cdot \sum_{\substack{i,j \in N \\ i \neq j}} \varepsilon_{ij} = \frac{1}{N \cdot (N-1)} \cdot \sum_{\substack{i,j \in N \\ i \neq j}} \frac{1}{d_{ij}}. \qquad (2)$$

The global efficiency is correlated to $1/L$, since a high characteristic path length corresponds to a low efficiency. By definition, in the topologic (non valued graph) case, $E_{glob}$ takes values in the interval [0,1], and is equal to 1 for the complete graph (a graph with all the possible, $N(N-1)/2$ edges). In metric systems (translated into valued graphs), however, it is possible to normalize (Latora and Marchiori, 2001) such a quantity dividing $E_{glob}(G)$ by the efficiency $E_{glob}(G^{ideal})$ of an ideal complete system in which the edge connecting the generic couple of nodes $i$-$j$ is present and has a length equal to the Euclidean distance between $i$ and $j$:

$$E_{glob}(G^{ideal}) = \frac{1}{N(N-1)} \sum_{i \neq j \in N} \frac{1}{d_{ij}^{Eucl}}, \qquad (3)$$

where $d_{ij}^{Eucl}$ is the Euclidean distance between nodes $i$ and $j$ along a straight line, i.e. the length of a virtual direct connection $i$-$j$.

A different normalization has been proposed in (Vragovìc et al. 2004):

$$E_{glob,2}(G) = \frac{\sum_{\substack{i,j \in N \\ i \neq j}} \frac{d_{ij}^{Eucl}}{d_{ij}}}{N(N-1)}, \qquad (4)$$

The three indices of centrality reported in (Freeman 1977, 1979) can be roughly divided into two different families (Latora e Marchiori, 2004). CD refers to being central as having a lot of people right at hand (a lot of "first neighbours"), while CC refers to being central as minimizing the distance to all other people in the group: thus, both can be seen as belonging to the same concept of being central as being near the others (Shimbel, 1953; Sabidussi, 1966, Nieminen, 1974; Freeman 1977, 1979; Scott, 2003). On the other hand, CB measures being central as being between the others, i.e. being the intermediary in many of the relationships that link all other persons each other in the group (Anthonisse, 1971; Freeman 1977, 1979; Freeman et al. 1991; Newman and Girvan, 2003). After a number of revisions and applications through a good four decades (Bonacich, 1972, 1987, 1991; Stephenson and Zelen, 1989; Altman, 1993), such indices have been changed and extended to different cases, but the basic families have not been changed as much. In transportation plan-ning, for instance, the accessibility of a place is still intended as its "ability" to be shortly accessed from all other places, which is in essence – beside distance being measured by a much more complex notion of transportation cost – a kind of CC.

The growth of interest for the network analysis of complex systems, an interest that has deeply involved physicists and mathematicians, new indices of centrality have been recently proposed. For the purposes of this article three of them, namely efficiency, straightness and information centrality, all based on the previously illustrated measure of global efficiency (Latora and Marchiori, 2001), are of particular relevance. The efficiency centrality $C^E$, which is in essence a closeness, once applied to properly *geographic* graphs and normalized by comparing the length of shortest paths with that of virtual straight lines between the same nodes (Vragovìc et al. 2004), turns out to capture a totally new, inherently geographic concept of centrality that we can term straightness centrality and is denoted as $C^S$: *being central as being more straightfully reachable by all others* in the network. The other index, namely the information centrality $C^I$ (Latora and Marchiori, 2004), in its two-steps computational process embeds both $C^C$ and $C^B$ in a single quantity, leading to another distinct concept of *being central as being critical for all others as a group*. In the following, we offer the formal mathematical definition and a more detailed discussion of the mentioned families of centrality indices.

*2.1 Being near the others: Degree and Closeness centrality*

Degree centrality is based on the idea that important nodes have the largest number of ties to other nodes in the graph. The *degree* of a node is, as previously mentioned, the number of edges incident with the node, i.e. the number of first neighbours of the node. The degree $k_i$ of node $i$ is defined in terms of the adjacency matrix as $k_i = \sum_{j \in N} a_{ij}$. The *degree centrality* ($C^D$) of $i$ is defined as (Nieminen, 1974; Freeman, 1979):

$$C_i^D = \frac{k_i}{N-1} = \frac{\sum_{j \in N} a_{ij}}{N-1}. \qquad (6)$$

The normalization adopted is such that $C^D$ takes on values between *0* and *1*, and is equal to one in the case in which a node is connected to all the other nodes of the graph. Degree centrality is not particularly relevant in primal urban networks where a node's degree (the number of streets incident in that node) are limited by geographic constraints.

The simplest notion of closeness is based on the concept of minimum distance or geodesic $d_{ij}$, that is, as already mentioned, the smallest sum of the edges lengths throughout all the possible paths in the graph from *i* to *j* in a weighted graph, and reduces to the minimum number of edges traversed, in a topologic graph. The closeness centrality of point *i* (Sabidussi, 1966; Freeman, 1979; Wasserman and Faust, 1994) is:

$$C_i^C = L_i^{-1} = \frac{N-1}{\sum_{\substack{j \in N \\ j \neq i}} d_{ij}} \qquad (7)$$

where $L_i$ is the average distance from actor *i* to all the other actors. $C^C$ is to be used when measures based upon independence are desired (Freeman, 1979). Such an index is meaningful for connected graphs only, unless one artificially assumes $d_{ij}$ equal to a finite value when there is no path between two nodes *i* and *j*, and takes on values between *0* and *1* in the case of non valued graphs.

*2.2 Being between the others: Betweenness centrality*

Interactions between two non-adjacent points might depend on the other actors, especially on those on the paths between the two. Therefore points in the middle can have a strategic control and influence on the others. The important idea at the base of this centrality index is that an actor is central if it lies between many of the other actors. This concept can be simply quantified by assuming that the communication travels just along geodesics. Namely, if $n_{jk}$ is the number of geodesics linking the two actors $j$ and $k$, and $n_{jk}(i)$ is the number of geodesics linking the two actors $j$ and $k$ that contain point $i$, the betweenness centrality of actor $i$ can be defined as (Freeman, 1979):

$$C_i^B = \frac{1}{(N-1)(N-2)} \cdot \sum_{\substack{j,k \in N \\ j \neq k; j,k \neq i}} \frac{n_{jk}(i)}{n_{jk}} \qquad (8)$$

$C_i^B$ takes on values between *0* and *1*, and it reaches its maximum when actor $i$ falls on all geodesics. There are several extensions to the index of betweenness proposed by Freeman. In particular, for all the cases in which the communication does not travel through geodesic paths only, a more realistic betweenness index should include non-geodesic as well as geodesic paths. Here we just mention two of the indices of betweenness that include contributions from non-geodesic paths: the flow betweenness and the random paths betweenness, although in our study we will only deal with the simplest case, i.e. the shortest-path betweeenness defined in formula (8).

*2.3 Being straight to the others: Efficiency and Straightness centrality*

Efficiency and straightness centralities originate from the idea that the efficiency in the communication between two nodes $i$ and $j$ is equal to the inverse of the shortest path lenght $d_{ij}$ (Latora and Marchiori, 2001). In particular, the efficiency centrality of node $i$ is defined as:

$$C_i^E = \frac{\sum_{\substack{j \in N \\ j \neq i}} \frac{1}{d_{ij}}}{\sum_{\substack{j \in N \\ j \neq i}} \frac{1}{d_{ij}^{Eucl}}}, \qquad (9)$$

where $d_{ij}^{Eucl}$ is the Euclidean distance between nodes $i$ and $j$ along a straight line.

The straightness centrality is a variant of the efficiency centrality, that originates from a different normalization (Vragovìc et al. 2004). The straightness centrality of node $i$ is defined as:

$$C_i^S = \frac{\sum_{\substack{j \in N \\ j \neq i}} \frac{d_{ij}^{Eucl}}{d_{ij}}}{N-1} \qquad (10)$$

This measure captures to which extent the connecting route between nodes $i$ and $j$ deviates from the virtual straight route.

*2.4 Being critical for all the others: Information centrality*

Information centrality is based on the idea that the importance of a node is related to the ability of the network to respond to the deactivation of that node from the network. In particular, we measure the network ability in propagating information among its points, before and after a certain node is deactivated.

The information centrality of a node *i* is defined as the relative drop in the network efficiency caused by the removal from *G* of the edges incident in *i*:

$$C_i^I = \frac{\Delta E_{glob,2}}{E_{glob,2}} = \frac{E_{glob,2}(G) - E_{glob,2}(G')}{E_{glob,2}(G)} \qquad (11)$$

where by *G'* we indicate a network with *N* points and $K-k_i$ edges obtained by removing from *G* the edges incident in the node *i*. Here we use the efficiency of Eq. 4. However, a generic performance parameter can be used in place of it. The removal of some of the edges affects the communication between various points of the graph increasing the length of the shortest paths. Consequently the efficiency of the new graph $E_{glob,2}(G')$ is smaller than $E_{glob,2}(G)$ and the index $C_i^I$ is normalized by definition to take values in the interval *[0,1]*. It is immediate to see that $C_i^I$ is somehow correlated to all the three standard centrality indices: $C_i^D$, $C_i^C$ and $C_i^B$. In fact, the information centrality of point *i* depends on the degree of point *i*, since the efficiency $E_{glob,2}(G')$ is smaller if the number $k_i$ of edges removed from the original graph is larger. $C_i^I$ is correlated to $C_i^C$ since the efficiency of a graph is connected to $L^{-1}_i$. Finally $C_i^I$, similarly to $C_i^B$, depends on the number of geodesics passing by *i*, but it also depends on the lengths of the new geodesics, the alternative paths that are used as communication channels, once the point *i* is deactivated. No information about the new shortest paths is contained in $C_i^B$, and in the other two standard indices.

*2.5 Space syntax in the field of centrality: integration and $C^C$*

The network approach has been broadly used in urban studies. Since the early sixties, a lot of research has been spent trying to link the allocation of land uses to population growth through lines of transportation (Wilson, 2000), or seeking the prediction of transportation flows given several topological and geometric characteristics of traffic channels (Larson, 1981), or eventually investigating the exchanges of goods and habits between settlements in the geographic space even in historical eras (Pitts, 1965, 1979; Peregrine, 1991; Byrd, 1994). Urban design as a discipline, beside some few theoretical efforts (Batty and Longley, 1994; Alexander, 1998; Salingaros, 1998) has not contributed that much to the picture in direct operational terms, with one quite relevant exception: after the seminal work of Hillier and Hanson (Hillier and Hanson, 1984), a rather consistent application of the network approach to cities, neighbourhoods, streets and even single buildings, has been developed under the notion of "space syntax", establishing a significant correlation between the topological accessibility of streets and phenomena as diverse as their popularity (pedestrian and vehicular flows), human way-finding, safety against micro-criminality, retail commerce vitality, activity separation and pollution (Penn and Turner, 2003). Though not limited to just one index, the core of the space syntax methodology, when applied to street networks, is the index of *integration,* which is stated to be «so fundamental that it is probably in itself the key to most aspects of human spatial organization» (Hillier, 1996, p.33). The integration of one street has been defined as the «shortest journey routes between each link (or a space) and all of the other in the network (defining 'shortest' in terms of fewest changes in direction)» (Hillier, 1998, p.36). As such, integration turns out to be nothing else than a normalized $C^C$ (Jiang and Claramunt, 2004b), the well known closeness centrality index defined since the early fifties by structural sociologists – who in fact dealt with topological networks as well – and reviewed by Freeman in the late Seventies. A short comparison between formal definitions of integration – Hillier and Hanson, (1984), ch.3, p.108; Teklenburg et al. (1993), p.350; Hillier, (1996), p.36, note 16; Jiang and Claramunt, (2002), p.298-299, and many others – and that of $C^C$ presented above in section 2.1, Eq. 7, fully confirms the assumption.

# 3 Primal vs. dual in the construction of 1-square-mile cases of urban street networks: metric distance, $C^C$ spatial distribution and the edge-effect

The network analysis, when applied to territorial cases, has mostly followed a primal approach, where intersections (or settlements) are turned into nodes and streets (or relationships) into edges. That representation seems to be the most intuitive for networks characterized by a strong connection to the geographic dimensions, which is to say networks where distance has to be measured not just in topological terms (steps) – like, for instance in social systems – but rather in properly spatial terms (meters, miles), like in urban street systems. Traffic engineers as well as economic geographers or even geo-archeologists have mostly, if not always, followed the primal approach. The primal approach is also the world-standard to geo-spatial datasets construction and diffusion, to the point that to date an immense amount of information has been already mar-keted following the embedded road-centerline-between-nodes rule, like in the huge TIGER (Topologically Integrated Geographic Encoding and Referencing) database developed at the U.S. Census Bureau; that rule, in fact, opens to the qualification of street/edges with weights of different kinds, i.e. traffic flows, transport costs, metric distance and the like.

It might appear paradoxical, though, that space syntax, the flagship application of urban design to the network analysis of city spaces, did follow the opposite direction, being based on a dual representation of urban street patterns. In this representation, axial lines that represent generalized streets (more exactly: "lines of sight" or "lines of unobstructed movement" along mapped streets) are turned into nodes, and intersections between each pair of axial lines into edges. Shortcomings as well as benefits of this approach have been often remarked (Desyllas and Duxbury, 2001; Jiang and Claramunt, 2002; Batty, 2004a,b; Ratti, 2004; Hillier and Penn, 2004; Porta et al. 2004). For the purposes of this article, we just focus on the questions of distance and on the impacts of the two approaches on the spatial distribution of $C^C$.

In the process of building the dual graph, which means reducing streets into nodes, what gets lost is something very relevant, beside its somehow questionable importance for the human *cognitive* experience of spaces (Penn, 2003), for any human *sensorial* experience of space (Hall, 1966): distance. No matter its real length, one street will be represented in the dual graph as one point. Moreover, as long as a generalization model is run and the "identity" of one real street is extended over a conceptually unlimited number of real intersections, in the dual graph it is possible to find one node (street) with a conceptually unlimited number of edges (intersections), a number which heavily depends on the actual length of the street itself. Thus, the longer the street in reality, the more central (by degree) it is likely to be in the dual graph, which counters the experiential concept of accessibility that is conversely related to how close (i.e. in real metric distance) is the destination to all origins, like in transportation models. Another consequence of the dual reduction of streets into points is that it makes impossible to account for the variations which so often characterize one single street, variations that may easily become very significant for lengthy streets that cross wide – and possibly quite different in structural terms – urban areas; this is the case, for instance, of the via Etnea in Catania, Sicily, a roughly 3 kilometres long perfectly straight XVII century street that runs from the baroque city core to the countryside beneath the Etna volcano, a street that presents radically different social, economic, demographic and environmental settings while crossing seemingly all possible urban landscapes on Earth.

Finally, in more structural terms, metric distance has been recognized as the key feature of road networks, which exactly because of that do need to be dealt with as a new, specific family of networks (Gastner and Newman, 2004); the crucial nature of geographic-Euclidean distance at the core of such systems leads to other key features, namely the planar nature and the extremely reduced variance of node's degree, whose distribution could never recall, and does not, any particular scale-free behavior. However, processed through the dual representation and a generalization model, the same road network gets freed of such limitations: the loss of any limitation to the degree of a node in the dual graph, which comes because the dual node represents one generalized real street and any generalized real street might cross a conceptually unlimited number of other real streets, makes the dual generalized streets systems structurally analogous to all other topological systems recently investigated in other fields, which in fact do not exhibit any geographic constrain; that leads, for instance, to the recognition of broad scaling rules in the degree distribution (Jiang and Claramunt, 2004a; Rosvall et al. 2004). The same scaling rules emerge for both the degree and the clustering coefficient $C(k)$ distributions of dually represented generalized streets networks of significant size (Porta et al, 2004), the sign of a hierarchical structure that has been recognized as a distinct property of non-geographic real cases, i.e. networks where the metric length of edges simply does not matter in any sense (Ravasz and Barabasi, 2003).

Hence, the dual representation associated with a generalization model – the two pillars of the space syntax castle – actually push urban street systems, in strict structural terms, out of the geographic domain. Though other means can be investigated in order to introduce the geographic distance on such a dual representation (Salheen and Forsyth, 2001, Salheen, 2003, Batty, 2004b), if a central role to geographic distance has to be recognized in a straight and plain manner, the primal road-centerline-between-nodes representation of street patterns, where intersections are nodes and streets are (weighted) edges, evidently appears the most valuable option.

The "metric problem" that so deeply differentiates the primal and the dual representations of street systems, heavily impacts on both the way to capture the behavior of centrality measures on networks and the way networks themselves are represented into graphs. Recently, we pioneered a systematic evaluation of the distributions of different centrality indices over eighteen 1-square-mile samples of urban fabrics drawn from a previous work of Allan Jacobs (Jacobs, 1993), in a pure geographic framework (Crucitti et al. 2005). Four of those cases (fig. 1), namely Ahmedabad (IN), Venice (IT), Richmond and Walnut Creek (CA), are hereby deepened in order to frame the opposition between the primal and the dual approach to the newtork analysis of street systems.

**Fig. 1**

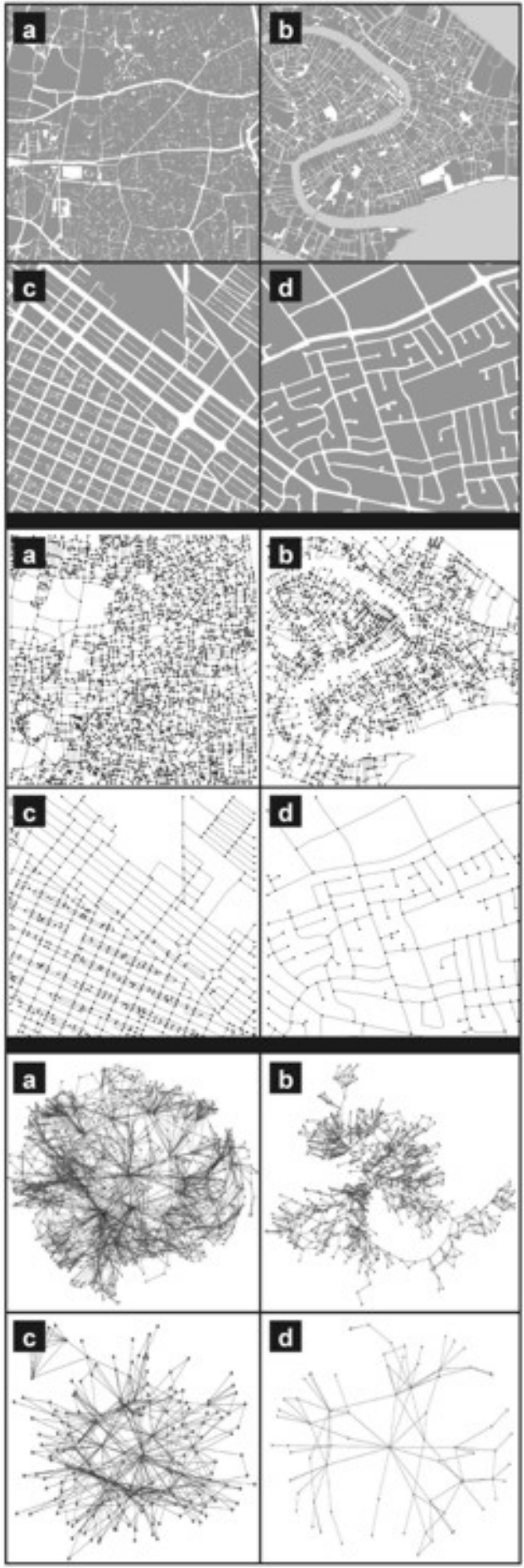

Four 1-square-mile cases of urban patterns as they appear in original maps (a-d, top), reduced to primal road-centreline-between-nodes graphs (a-d, middle), and dual generalized graphs (a-d, bottom). Two of them (a. Ahmedabad; b. Venice) are mostly self-organized patterns, while the others (c. Richmond; d. Walnut Creek) are mostly planned patterns. However, all cases are striking different after all other economic, historical, cultural, functional and geo-climatic conditions are considered. In particular, Ahmedabad is a densely interwoven, uninterrupted urban fabric while Venice is dominated by the Grand Canal separation which is over passed in just two points (the Rialto and Accademia bridges); moreover, Richmond shows a traditional grid-iron structure while Walnut Creek a conventional "lollipop" layout typical of post-war suburbs. Those geographic peculiarities, which are well featured in the primal valued (metric) representation, get lost in the dual, where just the topologic properties of the systems are retained.

Moreover, while Ahmedabad and Venice are typical self-organized patterns, in that they are mostly the outcome of a gradual, fine-grained development through history out of the control of any central agent, Richmond and Walnut Creek are examples of planned fabrics, developed under the directions of one central agent in a relatively short period of time.

In this work on 1-sq.mile samples of urban street systems (fig.1a-d, top) both the primal and the dual approaches have been performed and compared. In the primal approach centrality scores are calculated on nodes over the primal graphs. Primal graphs (fig. 1a-d, middle) are constructed following a road-centerline-between-nodes rule: real intersections are turned into graph nodes and real streets into graph edges; all graph edges are defined by two nodes (the endpoints of the arc) and, possibly, several vertexes (intermediate points of linear discontinuity); intersections among edges are always located at nodes; edges follow the footprint of real streets as they appear on maps. All distances have been calculated metrically in geographic space: so the distance between two nodes (intersections) is not – say – 3 steps (which is: through 3 intermediate nodes), but rather – say – 452.38 metres. After the calculation of centrality scores on primal nodes, analogous primal layouts (red/blue maps) are produced with reference either to node or edge centrality; in this latter case, because in the primal graph one edge is defined by just one pair of ending nodes by which the edge "participates" to the topology of the network as a whole, the centrality of one edge is simply equated to the average of the centralities of its defining pair of nodes. An example of node-referenced layout is given in fig. 2, while in fig. 3a an edge-referenced layout is offered: both are the result of the same pure primal approach.

In the dual approach centrality scores are calculated on nodes over the dual graph (fig.1a-d bottom) after the implementation of a generalization model, in this case the Intersection Continuity Negotiation (ICN) model (Porta et al. 2004), which emphasizes the straightness bias in collective wayfinding at intersections (Conroy Dalton, 2003; Thomson, 2004). Here, generalized streets are turned into nodes and intersections into edges of the dual graph; the distance between two nodes (streets) is equated to the number of intervening edges (intersections) along the shortest connecting path: it is a topologic, non-metric concept of distance which accounts for how many "steps" is one node from another, no matter the length of those steps. Then, primal layouts (red/blue maps) are produced where, because nodes in the dual graphs represent generalized streets, the centrality scores of a dual node are associated to the corresponding generalized street in the primal layout. Beside some dissimilarities, including a different generalization model (ICN instead of axial mapping), this follows the same approach of space syntax.

Before the analysis of different centrality indices, which is being presented in next sections, it is worth spending a little reflection on the sole closeness centrality $C^C$, the integration normalized index at the core of space syntax. Implemented on primal graphs (fig. 3a), the spatial distribution of $C^C$ is dominated by the so-called "edge-effect", in that higher $C^C$ scores consistently group around the geometric center of the image. To some extent less evident in less dense cases like Walnut Creek, the edge-effect is overwhelming in denser urban fabric like those of Ahmedabad and Venice. However, in all cases the edge-effect affects $C^C$ spatial distributions enough to disable the emergence of both central routes and focal spots in the city fabric, a crucial feature for urban analysis, so that it leads to meaningless results.

While the edge-effect dominates the primal representation, it is somehow minimized in the dual approach (fig. 3b), due to the combined impact of both the loss of metrics and, on the other side, the generalization model that makes streets less fragmented. Here we can see that the generalization model actually holds a vital role in that it allows to limit, to some extents, the edge-effect. How can it do that? Because by means of the generalization model the identification of continuous routes across the urban fabric is performed before the analysis of centrality rather than being one of its outcomes. As such, the analysis of centrality will give results that cannot avoid to be deeply affected by principles that do not belong to any concept of centrality, but instead belong to the algorithm embedded in the generalization model (straightness at intersections in the ICN, uninterrupted linearity – or visibility – in axial mapping). Thus our dual analysis, like that of space syntax, can be referred to as the combined result of two diverse and autonomous rationales, the first that drives the generalization process and the second that drives the spatial distribution of centrality. This finding confirms a previous work where a conventional space syntax dual analysis, applied without any generalization model on a segmentally represented street network, has been found misleading for the overwhelming impact of the edge-effect (Dalton et al. 2003). Again, this is not due to some hidden structure of the urban phenomenon: it is due to the inherent character of the chosen index: integration, or $C^C$, is quite a lot affected by the edge-effect due to its deep same nature, and does not lead by itself to any legible description of urban routes or focal areas.

In the light of this evidence, one has two options. The first option: you can insist with the dual approach and stress to its limits the empowerment of the generalization model, i.e. by means of more automation (Batty and Rana, 2002; Dalton et al. 2003) or new principles of street identity (Penn et al. 1997; Jiang and Claramunt, 2000; 2004b); Batty and Rana (2002) have found a good 9 different methodologies for axial mapping to date, each generating a different result from the others, all exhibiting substantially single scale length distributions. Also the edge effect can actually be minimized on dual graph by artificially widening the study area in order to leave the most edge-effected parts out of the picture, which seems a scarcely efficient – though truly effective – solution. The second option: you can embrace a pure primal approach, riveting everything to geographic metric distances weighted on a road-centerline-between-nodes world standard; this latter option would allow to reach a much finer characterization of (even longest) streets; it would let you abandon the generalization models, with a relevant advantage in process efficiency, objectivity and legibility; it also would open to endless, already marketed and constantly updated information resources; finally, it would lead to a great enhancement in the realism of calculations and representations, in that it pairs the *topology* of the system with its *metric*, thus comprehending both the cognitive and the proxemic dimension of collective behaviors in space. Striking benefits, as everyone can see. But there are one problem and one question.

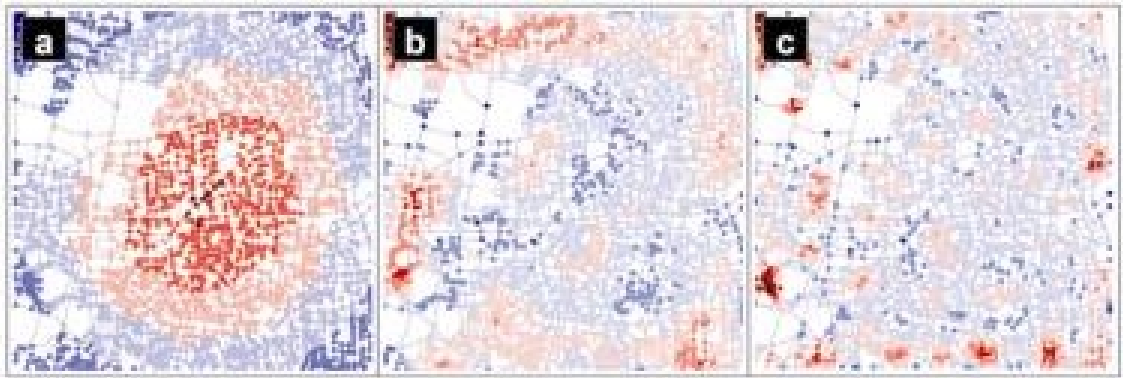

**Fig. 2**

The primal approach: $C^C$ spatial distribution in Ahmedabad: centrality scores are calculated on the nodes of the primal weighted graph (fig. 1a, middle) where weights are edges' metric lengths. In this figure red/blue maps have been anchored to nodes' centrality, while in others (fig. 3a) they have been referred to edges' centrality as the average centrality of its defining nodes. (a) Global Closeness: $C^C$ is calculated on the whole network; (b) local Closeness: $C^C$ is calculated on the sub-network of nodes at distance $d < 400$ meters from each node; (c) local Closeness: $C^C$ is calculated on the sub-network of nodes at distance $d < 200$ meters from each node. The $C^C$ index, performed globally over the whole network (see also fig. 3a, case 1), exhibits an overwhelming sensibility to the edge effect that makes it almost meaningless for urban analysis. However, performed locally (b) and (c), $C^C$ maintains a good descriptive potential, but it captures only local properties. The problem of the edge-effect with the global measure is partially corrected in the dual graph (fig. 3b, case 1) due to its topologic nature and the implementation of a generalisation model. However, beside just partially solving the problem, this correction seems artificial in that it cannot account for variations of centrality along lengthy streets; moreover, it is heavily dependent on the rationale of the generalization model, that has nothing to do with centrality.

The problem: as we have just shown, the $C^C$/integration index simply does not work on such primal graphs. It does not reveal any order in the graph, it just "reveals" the center of the image. But $C^C$ is not all: centrality is a multifold concept and we have many indices at hand. Thus, to overcome this problem we can limit the analysis of $C^C$ to a local scale, where it maintains a good potential (fig. 2), and simply begin with testing other centrality indices – like the mentioned $C^B$, $C^S$ and $C^I$ – that refer to different concepts of being central. That is what we basically do hereby; a review of our findings and a comparison with an application on dual graphs is offered in the next section. The question: is the primal approach to street systems, because of the inherent limitation of the $C^D$ range of variance to scores mostly included between 3 and 6 (roughly the number of streets per intersection in real urban patterns), getting the whole reflection unsuitable to the more general stream of the network analysis of complex real-world systems? Again, $C^D$ is not all as well: we will show that once the analysis of the statistical distribution of centrality over the network is extended from the sole $C^D$ to the other centrality indices, consistent scaling behaviors come to light that provide a much deeper insight in the complex nature of real street networks (and more in general of geographic systems). So, the evaluation of multiple centrality concepts and measures, it turns out, is the key for being able to both perform a pure primal road-centerline-between-nodes spatial analysis (with all its benefits) and, on the other side, re-enter the network analysis of geographic systems into the mainstream of the "new sciences of networks".

## 4 The spatial and statistical distributions of centrality indices

Differences and correlations among the many indices of centrality in social networks have been investigated in a significant flurry of literature over the last decades (Cook et al. 1983; Donninger, 1986; Bolland, 1988; Markovsky et al. 1988; Nakao, 1990; Mullen at al. 1991; Rothenberg et al. 1995; Bell et al. 1999; Poulin et al. 2000; see for a review: Wasserman and Faust, 1994); the goal was to understand the real nature of those indices when applied on human groups or organizations. The implementation of centrality indices on territory-related cases – though not always geographic – has been, to this respect, much less experimented, with some exceptions in different and sometimes unexpected fields (Pitts, 1965, 1979; Irwin-Williams, 1977; Irwin, 1978; Rothman, 1987; Peregrine, 1991; Byrd, 1994; Smith and Timberlake, 1995; Faust et al. 1999). In space syntax, for instance, the link between the $C^C$/integration core-index (as well as the ancillary $C^D$/connectivity index) and centralities in social networks has been only very recently acknowledged (Jiang and Claramunt, 2004b) and never really deepened, thus there is apparently no comparative evaluation with other families of centrality indices. Our studies on 1-square-mile samples of urban street networks applied in a pure primal, geographic framework (Crucitti et al. 2005) reveal that the four families of centrality, "being near" ($C^C$), "being between" ($C^B$), "being straight" ($C^S$) and "being critical" ($C^I$) exhibit highly diverse spatial distribution patterns (fig. 3a). In this sense, two conclusions occur: on one hand, no single index gives the whole picture, as they tell striking different stories: in fact, they capture different aspects of one place's "being central" in geographic space; on the other hand, $C^I$ emerges as the most comprehensive single index of the whole set, gathering properties of two different families of centrality out of the remaining three that we have taken into consideration. More in detail, $C^C$ (performed globally) fails in individuating a hierarchy of central routes or areas and is of no help in urban analysis; $C^B$ is mostly effective in letting centrality emerge along even lengthy urban routes while still being affected, to some extents, by the edge-effect; $C^S$ gives the most unexpected results, mapping areas of higher centrality as well as central routes, with no sensibility to the edge-effect; $C^I$ nicely capture the criticality of edges that play a "bridging" role in keeping the network connected, at the same time partially retaining the behavior of $C^B$. In general, the particular effectiveness of the analysis to account for the variations in centrality levels within the same route, like in the case of $C^B$ in Venice, $C^S$ in Ahmedabad or $C^I$ in Richmond, should be highlighted: even more so if considered that those routes are a result of the "natural" emergence of a pattern of centrality across a plurality of street segments, without any hexogen intervention of rationales of a different kind like that of a generalization model: as such, MCA suggests that centrality can play a distinct role in the "organic" formation of a "skeleton" of mostly practiced routes as the cognitive framework for wayfinding in a complex urban environment (Kuipers et al. 2003).

In fig. 3b an analogous assessment over a dual representation of the same cases is presented, after the implementation of the ICN generalization model (Porta et al. 2004). The analysis of $C^C$ gives a good result (see for instance the case of Ahmedabad) but still affected by the edge-effect as the street pattern, like in Venice, gets more fragmented. The ICN-driven construction of generalized streets, which is preliminary to the process of centrality calculation, deeply affects the final results in all cases, leading to a more artificial picture of real systems and to a less differentiated information among indices. While in the primal approach continuous routes or sub-areas emerge in the urban fabric as a result of the natural "convergence" of centrality over a chain of single streets across a number of intersections, in the dual we have routes that are identified before centrality enters the scene, and *then* attributed a value of centrality. That leads to less univocal results, in that it is impossible to understand whether a certain part of a route is really as central or it just "inherits" its centrality from others of its parts. All this turns out in a sort of artificial interpretation of the street network.

Moreover, the statistical analysis of centrality distributions on primal graphs (fig. 4, top) confirms that the cumulative distributions of $C^C$, $C^B$, $C^S$ and $C^I$ consistently follow characteristic scaling rules: here, like in all samples considered in Crucitti et al. (2005), we have found that $C^C$ and $C^S$ are mainly linear; $C^B$, being well fitted by a straight line in a log-linear scale, follows an exponential distribution; $C^I$ differentiates planned from self-organized cities following an exponential distribution in the former case and a power-law in the latter one. In fig. 4a we report the distribution of centrality indices for Ahmedabad and Richmond, representative examples of self-organized and planned cities respectively. Homogeneity/heterogeneity in the allocation of the centrality "resource" among nodes, investigated by the application of the Gini coefficient (Dagum, 1980), have demonstrated to be sufficient for a broad classification of different cities through a cluster analysis, that groups together cities with similar urban patterns (Crucitti et al. 2005). This confirms that, by means of the primal representation and a set of different centrality indices, it is possible to capture basic crucial properties of real urban streets systems for an appropriate classification of cities.

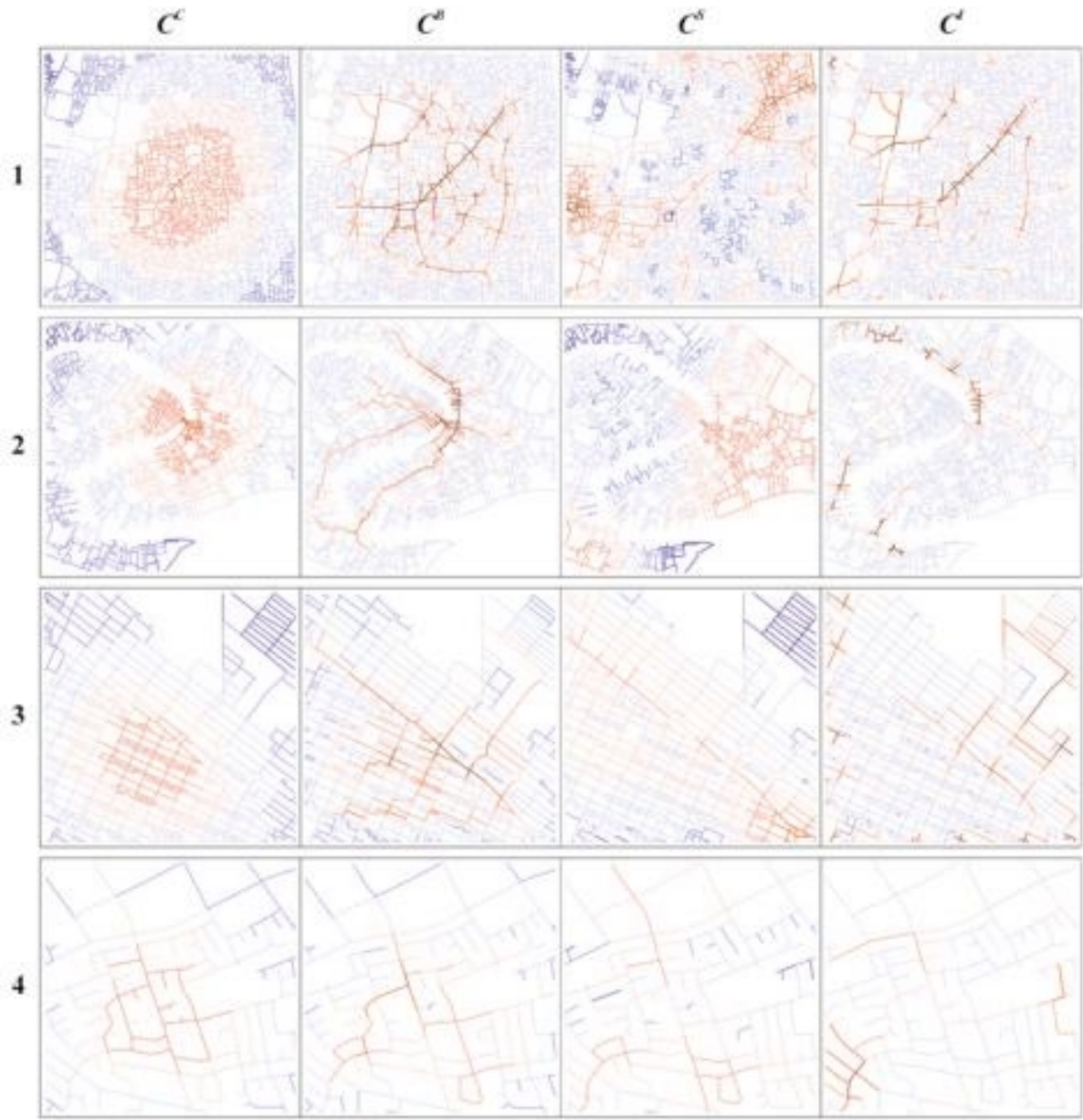

**Fig. 3a**

The primal approach: the four indices of centrality ($C^C$=Closeness; $C^B$=Betweenness; $C^S$=Straightness and $C^I$=Information) calculated over four cases (1. Ahmedabad; 2. Venice; 3. Richmond; 4. Wallnut Creek) represented by primal graphs (intersections are nodes, streets are edges). In the primal approach, centrality scores are calculated metrically on nodes, that represent real intersections; then, they can be represented in red/blue layouts as referred to nodes or, like in this figure, to edges: in this latter case, each edge is assigned a centrality score equal to the average score of its pair of nodes, through which the edge participates to the topology of the whole network. The analysis shows that $C^C$ is by far the less interesting measure of centrality, lacking any capacity to identify central routes or sub-areas in the urban pattern due to the overwhelming impact of the edge-effect. Conversely, the three other indices consistently capture the different "natures" of being central. Moreover, they exhibit a strong capacity to identify routes or sub-areas of higher centrality in the most diverse topologic conditions. Thus, that identification is an outcome of the sole centrality analysis: in short, we see here how centrality "builds" different routes or spots in the urban fabric, whose identity is a function of their qualification in terms, again, of "pure" centrality.

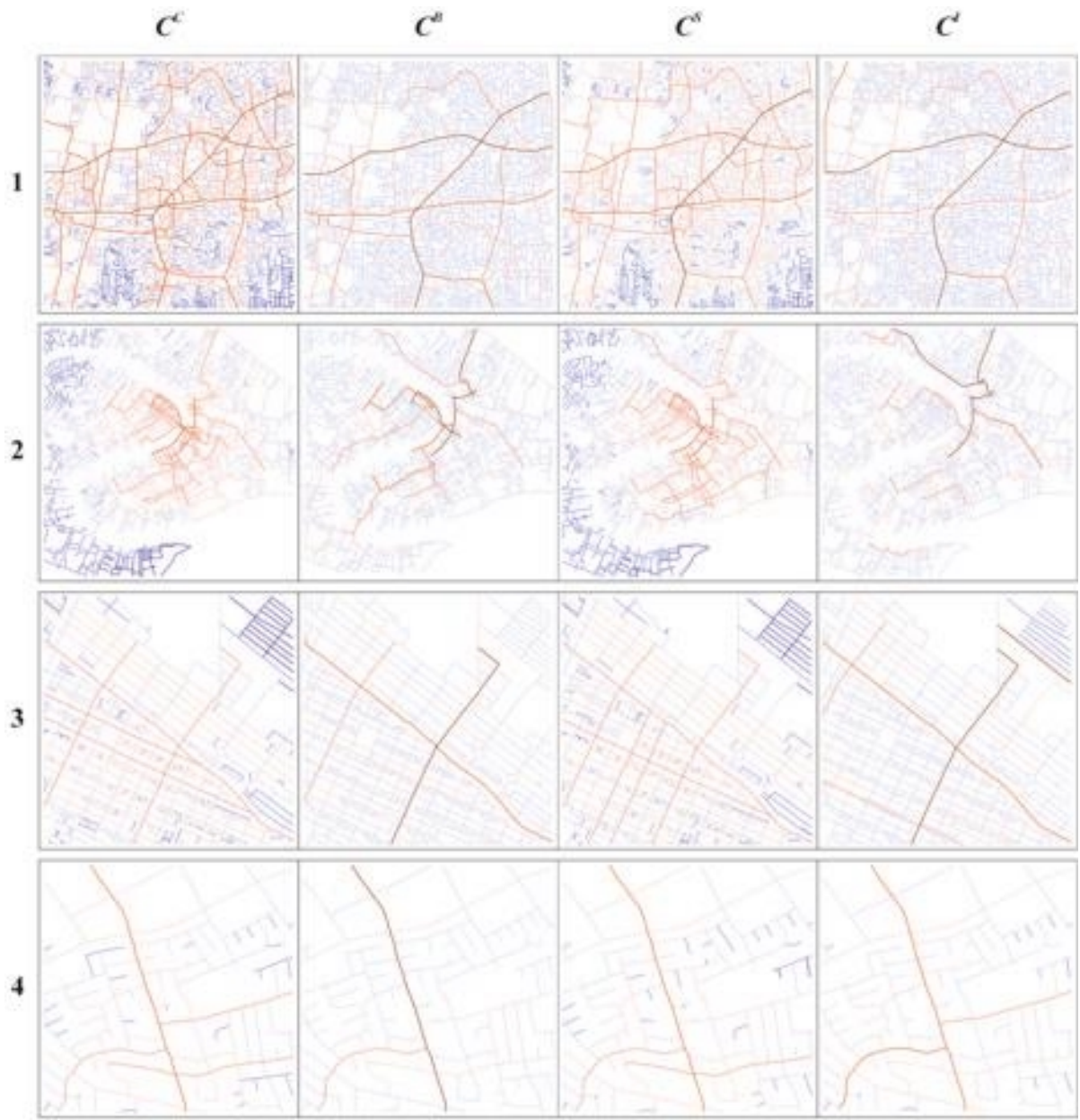

**Fig. 3b**

The dual approach: the four indices of centrality ($C^C$=Closeness; $C^B$=Betweenness; $C^S$=Straightness and $C^I$=Information) calculated over four cases (1. Ahmedabad; 2. Venice; 3. Richmond; 4. Wallnut Creek) represented by dual graphs (generalized streets are nodes, intersections are edges). In the dual generalized approach centrality scores are calculated topologically on nodes, that represent generalized streets (single-identity routes formed by a plurality of sequential real streets). The passage from a set of streets to a set of generalized streets is obtained by the implementation of a generalization model. In the figure, we have used the previously defined Intersection Continuity Negotiation (ICN) model (Porta et al. 2004), which is based on straightness at intersections; in conventional space syntax, the axial mapping process embeds a generalization model based on uninterrupted linearity; variations have been recently proposed to axial mapping based on a concept of fractional depth (Dalton, 2001; Dalton et al. 2003), characteristic points of navigation decision (Jiang and Claramunt, 2000) or named streets (Jiang and Claramunt, 2004b); in any case, generalized streets are defined before the implementation of the centrality analysis, thus introducing a rationale that has nothing to do with centrality in itself, while deeply affecting the results of the whole process. Thus, centrality does not contributes to the identification of the urban routes that we see in red/blue maps above: it just qualifies them after they have been identified.

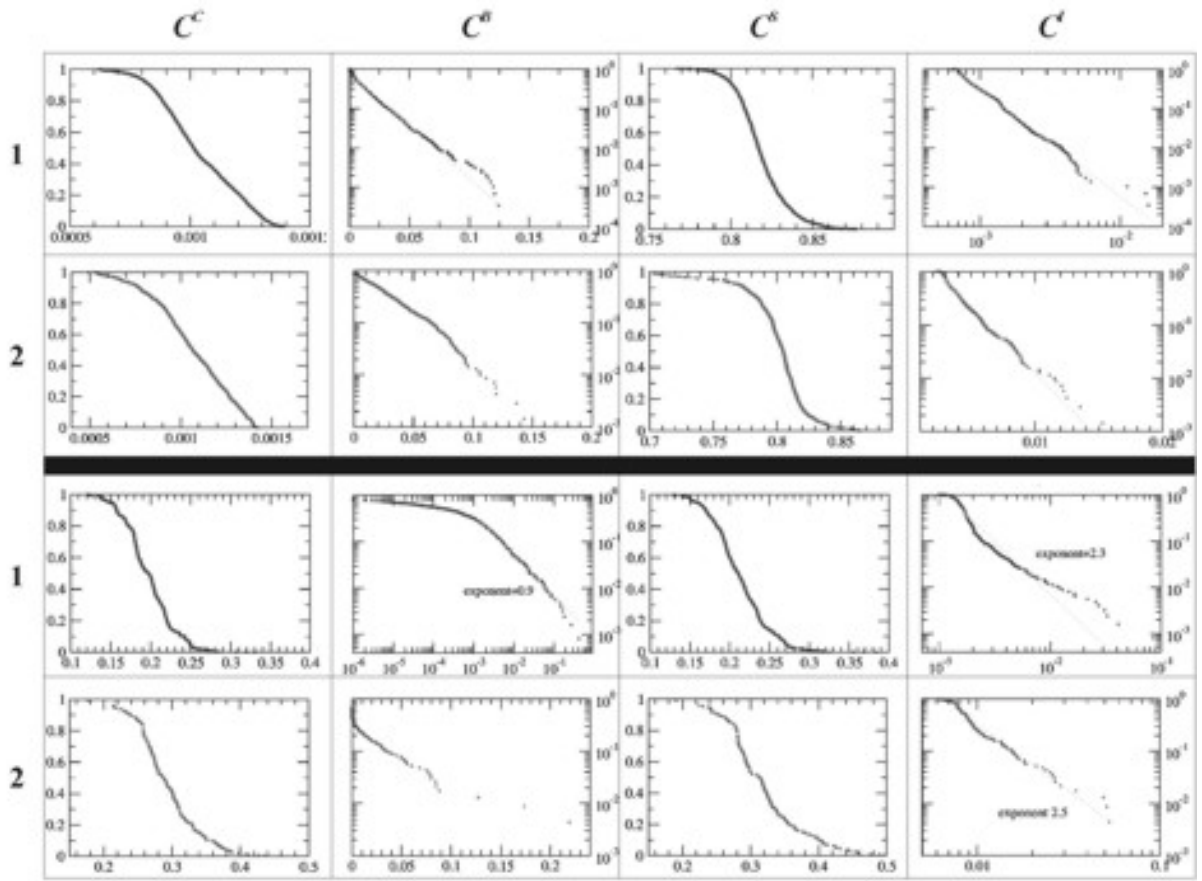

**Fig. 4**

The cumulative distribution of the four indices of centrality ($C^C$=Closeness; $C^B$=Betweenness; $C^S$=Straightness; $C^I$=Information) in the primal (top) and in the dual (bottom) graph representations of Ahmedabad (case 1) and Richmond (case 2). Cumulative centrality distributions P(C) are defined by $P(C) = \int_C^\infty N(C')/N \, dC'$, where N(C) is the number of nodes having centrality equal to C.

The primal representations acknowledge the existence of consistent behaviours of the same index across all cases, while different behaviours emerge for different indices: $C^C$ and $C^S$ are mainly linear; $C^B$ *is* single scale and the dashed lines in the linear-logarithmic plot show an exponential distribution P(C)~exp(-C/s) for self-organized cities (like Ahmedabad, $s_{Ahm}$ = 0.016), and a Gaussian distribution P(C) ~exp(-1/2 $x^2/\sigma^2$) for mainly planned ones (like Richmond, $\sigma_{Rich}$=0.049). Such distinction is more emphasized in $C^I$ that follows a power law distribution P(C)~$C^{-\gamma}$ in self-organized cities ($\gamma_{Ahm}$=2.74), and an exponential distribution in planned cases ($s_{Rich}$=0.002).

The dual representations: $C^C$ and $C^S$ are mainly linear. $C^B$ and $C^I$ exhibit broad-scale and single-scale distributions respectively for Ahmedabad and Richmond. Although it is possible to intuit the kind of law that would better represent each distribution, the deviation from the analytic curves is often very large.

Differently than in primal, in dual graphs the statistical distribution of $C^D$ is a relevant feature because the number of intersections per street is conceptually unlimited (see section 3); addressed in a previous work (Porta et al. 2004), the $C^D$ distribution was found to follow a power-law rule in systems of a relevant size. The same analysis is hereby extended to the other centrality indices (fig. 4, bottom), revealing that, as in the primal representation, also in the dual $C^C$ and $C^S$ are mainly linear, $C^B$ and $C^I$ seem to follow broad-scale distributions for Ahmedabad and single-scale distributions for Richmond. Nevertheless, the distributions are much less clear than in the primal analysis (notice for instance the deviation from the analytic fit in the $C^I$ of Ahmedabad), confirming again that the primal representation has greater capabilities to extract such hidden order from urban patterns.

# 5 Conclusions: the benefits of the primal approach and the Multiple Centrality Assessment

A network analysis of four 1-square-mile samples of urban street systems is performed over primal and dual graphs. This introduces some considerations that strongly support the primal approach as a more comprehensive, objective, and realistic methodology for the network analysis of centrality on urban street systems. Being based on the world standard road-centreline-between-nodes format, the primal approach is suitable for making the best use of huge information resources developed and available in a broad variety of different fields. Moreover, the primal approach significantly reduces subjectivism in graph construction by excluding problems related to the generalization model. While in the dual approach centrality statistical distributions, beside the case of $C^D$ in systems of relevant size, exhibit curves that significantly deviate from analytical fits, clear rules evidently and consistently emerge in the primal representation: this alone witnesses the higher power of the primal approach to capture some basic crucial structures of real urban streets systems. That seems to be inherently linked to the by far most relevant difference between the two options: while the primal representation allows to preserve a metric, geographic concept of distance, without abandoning the topology of the system, the dual approach necessarily leads to just a topological step-distance concept that makes indices and processes fundamentally more abstract, in that they miss much of the sensorial dimension of human ecology. Finally, our work shows that centrality is not just one single thing. Centrality is a multifaceted concept that, in order to measure the "importance" of single actors, organizations or places in complex networks, has led to a number of different indices. We show that such indices, at least those mentioned in this article, do actually belong to four deeply different concepts of being central as being near, being between, being straight to and being critical for the others in a geographic system: the diversity of these "families" is witnessed by the consistently different distribution of centrality scores in considered cases, both in terms of *spatial* distributions as mapped in red/blue layouts and in terms of *statistical* distributions as mapped in cumulative plots. We also show that such indices, when applied to geographic networks, capture different ways for a place to be central, each way quite significant for a deeper understanding of how cities work. A good overview of such heterogeneity of being central in a street network is offered in fig. 3a, where betweenness, straightness and information centrality "describe" the same city differently while exhibiting a good consistency across cases.

A new approach to the network analysis of centralities in geographic systems, such as urban street systems, is therefore appearing. Its three pillars are: 1. the use of primal graphs; 2. the use of metric distance; 3. the use of many different indices of centrality. As such, we may well name it Multiple Centrality Assessment (MCA). Advantages of MCA, if compared to dual approaches like space syntax, are multifold: 1. it is not based on any generalization model, therefore is more legible, feasible and objective; i.e. it leads to the identification of central routes by the sole "natural" convergence of centrality over chains of segmented streets; 2. it is fit to access the huge amount of information resources developed under the road-centreline-between-nodes world standard, including network constructed for traffic engineering and modelling or geo-mapping worldwide; 3. it is more comprehensive and realistic, in that it couples the *topology* with the *metrics* of the system; 4. it gives a set of multifaceted pictures of reality, rather than just one: that leads to more argumentative, thus less assertive, indications for action. Moreover, from a pure research perspective, MCA makes it possible to analyse geographic networks under the same roof of well established studies of non-geographic networks such as biologic, technological or social networks.

On this basis, further research may well proceed in three directions: on one hand, procedures for the systematic evaluation of proposed indices should be coordinated and standardized under a single tool; on the other hand, significant achievements are likely to be gained after establishing correlations between centralities *of* the networks (which are structural measures) and dynamics *on* the networks (like patterns of land-uses, real-estate values, social groups allocation, crime rates, retail allocation, pedestrian and vehicular flows and others), i.e. by the re-coding of centrality indices in the new context of "weighted" networks (Barrat et al. 2004). It should not be of surprise in fact, as a general hypothesis, if some dynamics turn out to be strictly correlated to one family of centrality indices while others are more correlated to a different family, which would tell a lot about the deep meanings of being central for a place in real geographic systems and would open to a more targeted use of the tool in urban planning and design. Finally, a specific effort should be produced to normalize all centrality measures in order to make the comparison between systems of different size possible.